\documentclass[12pt]{article}
\usepackage{amsmath}
\usepackage{amsfonts}
\usepackage[round]{natbib}
\usepackage{hyperref}
\usepackage{authblk} % To add affiliation
\usepackage{multirow} % For multi-row cells in tables
\usepackage{array} % For defining custom column width
\usepackage{geometry} % For adjusting page layout 
\usepackage{pdflscape}
\usepackage{caption} % For customizing the caption format
\usepackage{placeins}
\hypersetup{
    colorlinks=true,
    linkcolor=blue,
    citecolor=blue,
    urlcolor=blue
}

\title{MICG-AI: A multidimensional index of child growth based on digital phenotyping with Bayesian artificial intelligence}
\author[*,$\dag$]{Rolando Gonzales Martinez}
\author[*]{Hinke Haisma}
\affil[*]{\small University of Groningen, Faculty of Spatial Sciences}
\affil[$\dag$]{University of Oxford, Oxford Poverty and Human Development Initiative}

\date{December 2024}

\begin{document}

\maketitle

\begin{abstract}
\noindent This document proposes an algorithm for a mobile application designed to monitor multidimensional child growth through digital phenotyping. Digital phenotyping offers a unique opportunity to collect and analyze high-frequency data in real time, capturing behavioral, psychological, and physiological states of children in naturalistic settings. Traditional models of child growth primarily focus on physical metrics, often overlooking multidimensional aspects such as emotional, social, and cognitive development. In this paper, we introduce a Bayesian artificial intelligence (AI) algorithm that leverages digital phenotyping to create a Multidimensional Index of Child Growth (MICG). This index integrates data from various dimensions of child development, including physical, emotional, cognitive, and environmental factors. By incorporating probabilistic modeling, the proposed algorithm dynamically updates its learning based on data collected by the mobile app used by mothers and children. The app also infers uncertainty from response times, adjusting the importance of each dimension of child growth accordingly. Our contribution applies state-of-the-art technology to track multidimensional child development, enabling families and healthcare providers to make more informed decisions in real time.
\end{abstract}

\newpage

\section{Introduction}

Digital phenotyping is the use of digital devices, especially smartphones and mobile applications, to collect real-time, high-frequency data that can reveal an individual’s behavioral, psychological, and physiological state in naturalistic settings. By passively or actively gathering data on user interactions, location, motion, and response patterns, digital phenotyping enables researchers to capture fine-grained and ecologically valid behavioral data that may not be accessible through traditional survey methods \citep{Torous2018, Huckvale2019, Insel2017}. This approach is especially relevant in behavioral health, where response times, engagement levels, and even app usage patterns provide insights into emotional or cognitive states. Digital phenotyping has been applied to track mood and monitor general well-being by analyzing subtle shifts in behavioral data captured continuously over time \citep{Jain2015, Onnela2016}. Furthermore, it offers a scalable, non-invasive means of behavioral monitoring, allowing researchers to gather data across diverse populations in real-world contexts.

In the context of child development, understanding multidimensional growth requires moving beyond traditional physical metrics to include psychological, social, and environmental dimensions. Multidimensional child growth encompasses a range of developmental dimensions such as cognitive, emotional, and social and environmental contexts, in addition to physical growth. Research increasingly highlights the need to track multiple dimensions of development to capture a holistic picture of a child’s well-being and developmental trajectory \citep{Black2017, Richter2017, Britto2017}. Comprehensive models of child growth take into account not only the biological and physical dimensions but also the psychosocial and environmental contexts that shape a child’s development. Previously, a Multidimensional Index of Child Growth (MICG) that assesses various dimensions of child development beyond traditional physical growth metrics was proposed by \citet{gonzales2022}, based on the conceptual foundations of the The Task Force ``Towards a Multi-dimensional Index to 14 Child Growth'' of the International Union of Nutritional Sciences \citep{haisma2018towards}. This approach, grounded in the capability framework, includes social, cognitive, environmental, and emotional indicators to provide a comprehensive assessment of a child's well-being and development \citep{Yousefzadeh2018}. The MICG model, however, is challenging to implement in real-world settings, especially at scale, due to the intensive data collection and analysis they require. Integrating multidimensional indicators in a single framework remains a complex task due to the variability and interconnected nature of these growth dimensions \citep{Grantham2016, Walker2007}.

The purpose of this paper is to propose a Bayesian artificial intelligence algorithm that applies digital phenotyping to track multidimensional child growth in real time. By embedding this algorithm within a mobile application provided to mothers and children, we aim to create a user-friendly tool for monitoring the developmental progress of children across multiple dimensions. The Bayesian approach enables the model to dynamically update and refine predictions based on new data inputs, capturing both immediate behavioral data and the subjective opinions of mothers and children regarding growth indicators. This probabilistic framework allows for the handling of uncertainty and variability inherent in subjective and behavioral data. Furthermore, by using digital phenotyping, the app can weigh responses based on inferred certainty from response patterns, creating a more accurate and comprehensive index of child growth that adapts over time. Through this approach, we aim to offer a scalable, practical tool for families and healthcare providers to gain continuous insights into child development.

The following section provides background information on the theory and outline potential ways to aggregate individual capabilities into broader dimensions of multidimensional child growth. Section \ref{sec:Bay} introduces a Bayesian approach for calculating the weights of each indicator within these growth dimensions. Section \ref{sec:ga} presents an algorithm designed to optimize digital phenotypes by adjusting weights using genetic algorithms and a neural network-based fitness function. Finally, Section \ref{sec:con} concludes.

\section{Aggregating Capabilities into Dimensions of Multidimensional Child Growth}

This section presents a hierarchical aggregation of the MICG indicators, progressively grouping them into broader dimensions for a holistic understanding of child development. Table \ref{table:MICG} shows the 12 capabilities and the indicators previously used to calculate the MICG in practice by \citet{gonzales2022}, based on the 12 dimensions of child growth of \citet{biggeri2006children} and the theoretical capability approach to multidimensional child growth developed by \citet{haisma2018towards} and \citet{Yousefzadeh2018}. Table \ref{table:MICG} is presented as an example of how the MICG can be calculated in practice. The indicators, however, can be modified based on data availability, contextual factors, and the objectives of digital phenotyping.

%\begin{landscape} % Rotate the entire page to landscape orientation
\begin{table}[ht]
\centering
\small % Reduce font size of the table
\caption{Indicators included in the multidimensional index of child growth}
\label{table:MICG}
\resizebox{\textwidth}{!}{%
\begin{tabular}{l l p{10cm}} % Remove vertical lines by not using "|"
\hline
\textbf{Capability} & \textbf{Indicator} & \textbf{Deprivations cutoff for the indicators*} \\ \hline

\multirow{6}{*}{Life and physical health} 
& overweight & Child deprived if BMI \( > \) 2SD (WHO thresholds depending on the age/sex of the child) \\ 
& stunting & Deprived (stunted) if child has a length/height-for-age z-score of less than -2SD \\ 
& wasting & Deprived (wasted) if child has a weight-for-length/height z-score of less than -2SD \\ 
& nutrition (frequency) & Deprived if child eats less than 4 times per day \\ 
& nutrition (diversity) & Deprived if child eats less than 4 different food groups per day \\ 
& vaccination & Deprived if child has not a vaccination card \\ \hline

\multirow{3}{*}{Bodily integrity and safety} 
& health & Deprived if child has worse health than other children \\ 
& safety in the street & Deprived if parents think it is not safe for children to go on the street on her/his own \\ 
& shelter & Deprived if child lives in a household made of rudimentary material (roof/wall): straw, wood, leaves, mud, plastic sheets \\ 
& danger & Deprived if child feels/would feel in danger when travelling to school \\ \hline

\multirow{4}{*}{Love and care} 
& love of parents & Deprived if parents believe that love between a parent and child is not important at all or not very important \\ 
& proud of children & Deprived if parents strongly disagree about feeling proud of their children \\ 
& parents' responsibility & Deprived if parents believe that sense of responsibility is not important at all or not very important \\ 
& parents' fulfillment & Deprived if the pleasure parents get from watching children grow is not important at all or not very important for them \\ \hline

\multirow{2}{*}{Leisure activities} 
& leisure & Deprived if child does not spend time with friends/younger children, playing games, watching TV, playing alone or with pets \\ \hline

\multirow{1}{*}{Respect} 
& respect & Deprived if parents believe that it is not important to learn (i) responsibility, (ii) obedience, or (iii) respect for elders \\ \hline

\multirow{1}{*}{Social relations} 
& friendship & Deprived if child has difficulties making friends or cannot make friends at all \\ \hline

\multirow{1}{*}{Participation} 
& cooperation/participation & Deprived if parents think that it is no important to learn cooperation or participation at home \\ \hline

\multirow{2}{*}{Mental wellbeing} 
& cognitive & Deprived if child has a Rasch cognitive development score below the first quintile \\ 
& verbal & Deprived if child has a Rasch verbal PPVT score below the first quintile \\ \hline

\multirow{2}{*}{Education} 
& access to education & Deprived if the household does not have access to education \\ 
& imagination & Deprived if parents believe that it is not important to learn imagination at home \\ \hline

\multirow{1}{*}{Economic freedom} 
& paid/unpaid work & Deprived if child performs paid activities or if the child works on family farm or business on a typical day \\ \hline

\multirow{2}{*}{Environment} 
& external risks & Deprived if child is exposed to natural hazards or harassment from other children, authorities or from rebels/thieves \\ 
& internal risks & Deprived if number of children born before to same mother are higher than the ideal number according to caregivers \\ \hline

\multirow{2}{*}{Religion and identity} 
& children's religion & Deprived if she/he has a minority religion \\ 
& children's ethnicity & Deprived if she/he belongs to an ethnic minority \\ \hline

\multirow{1}{*}{Time autonomy} 
& domestic tasks & Deprived if she/he cares for others or performs domestic tasks on a typical day \\ \hline

\multirow{1}{*}{Mobility} 
& time to get to school & Deprived if she/he needs/would need more than one hour to get to school \\ \hline

\end{tabular}%
}
\end{table}
%\end{landscape}

\FloatBarrier

The MICG capabilities can be grouped into six primary dimensions:

\begin{enumerate}
    \item \textbf{Physical Health and Nutrition}: Overweight, stunting, wasting, nutrition (frequency), nutrition (diversity), vaccination.
    \item \textbf{Safety and Security}: Safety in the street, shelter, danger.
    \item \textbf{Emotional and Social Well-being}: Love of parents, pride in children, parents' responsibility, parents' fulfillment, friendship, respect.
    \item \textbf{Leisure and Learning Opportunities}: Leisure, cognitive development, verbal skills, access to education, imagination.
    \item \textbf{Economic and Environmental Stability}: Paid/unpaid work, external risks, internal risks.
    \item \textbf{Cultural and Personal Identity}: Children's religion, identity, autonomy in tasks, mobility to school.
\end{enumerate}

These constructs can be further condensed into four broader dimensions:

\begin{enumerate}
    \item \textbf{Health and Safety}: Combining physical health, nutrition, safety, and security indicators.
    \item \textbf{Emotional and Social Development}: Emotional and social well-being indicators, leisure, and social relations.
    \item \textbf{Learning and Cognitive Development}: Cognitive, verbal, educational access, and imaginative capabilities.
    \item \textbf{Economic, Environmental, and Identity Stability}: Economic freedom, environmental stability, religion, identity, autonomy, and mobility.
\end{enumerate}

Finally, the constructs of MICG can be condensed into three overarching dimensions:

\begin{enumerate}
    \item \textbf{Physical and Environmental Well-being}: Health, safety, and environmental stability indicators.
    \item \textbf{Emotional and Social Development}: Emotional well-being, social relations, identity, and cultural factors.
    \item \textbf{Educational and Cognitive Development}: Learning opportunities, cognitive and verbal skills, economic freedom, autonomy, and mobility.
\end{enumerate}

This hierarchical structuring enables a flexible analysis of multidimensional child growth, facilitating a nuanced, weighted aggregation across various levels of well-being. Formally, let the $12$ capabilities be:
\[
k_1, k_2, \ldots, k_{12}.
\]
In each $i$-indicator of these capabilities $x_i$, \( x_i = 1 \) represents a deprivation, and \( x_i = 0 \) represents non-deprivation.

Each indicator \(x_i\) is assigned a weight \(\omega_i\) such that \(\sum_{x \in C_i} \omega_i = 1\) for each capability \(k\).

The aggregation into 6 dimensions will be:
\begin{align*}
    C_1 &= \{k_1, k_2\} & \text{(Physical Health and Nutrition)} \\
    C_2 &= \{k_3, k_4\} & \text{(Safety and Security)} \\
    C_3 &= \{k_5, k_6\} & \text{(Emotional and Social Well-being)} \\
    C_4 &= \{k_7, k_8\} & \text{(Leisure and Learning Opportunities)} \\
    C_5 &= \{k_9, k_{10}\} & \text{(Economic and Environmental Stability)} \\
    C_6 &= \{k_{11}, k_{12}\} & \text{(Cultural and Personal Identity)}
\end{align*}

If a construct \( D_i \) is calculated as the weighted sum of $x_i$ indicators:
\[
D_i = \sum_{x \in C_i} \omega_i x
\]
where \(\sum_{x \in C_i} \omega_i = 1\), an aggregation into 4 broader dimensions will be:
\begin{align*}
    G_1 &= \{D_1, D_2\} & \text{(Health and Safety)} \\
    G_2 &= \{D_3\} & \text{(Emotional and Social Development)} \\
    G_3 &= \{D_4\} & \text{(Learning and Cognitive Development)} \\
    G_4 &= \{D_5, D_6\} & \text{(Economic, Environmental, and Identity Stability)}
\end{align*}

where each broader dimension \( G_i \) is calculated as:
\[
G_i = \sum_{D \in G_i} \omega_{G_i} D
\]
and \(\sum_{D \in G_i} \omega_{G_i} = 1\).

Finally, reducing the indicators to 3 overarching categories will be:
\begin{align*}
    H_1 &= \{G_1, G_4\} & \text{(Physical and Environmental Well-being)} \\
    H_2 &= \{G_2\} & \text{(Emotional and Social Development)} \\
    H_3 &= \{G_3\} & \text{(Educational and Cognitive Development)}
\end{align*}

The aggregated scores for each dimension \(H_i\) are:
\[
H_i = \sum_{G \in H_i} \omega_{H_i} G
\]
where \(\sum_{G \in H_i} \omega_{H_i} = 1\).

This hierarchical, weighted aggregation of binary indicators into constructs and dimensions enables a nuanced, weighted assessment of child growth across multiple dimensions of well-being.

\section{Bayesian approach to calculating the weights}\label{sec:Bay}

\subsection{Prior Distribution}

Given a data matrix \( X \in \mathbb{R}^{n \times k} \) with \( n \) children and \( k \) binary indicators, each entry \( x_{ij} \) represents the indicator for child \( i \) on dimension \( j \), a Gaussian prior distribution for each weight \( \omega_j \) will be:
\[
\omega_j \sim \mathcal{N}(\mu_{\omega_j}, \sigma_{\omega_j}^2)
\]
To establish a Gaussian prior distribution for each weight \( \omega_j \) using Likert-scale responses from children and mothers, the following steps can be followed:

\begin{enumerate}
    \item Likert Scale Responses: Each respondent provides two ratings for each indicator \( x_j \):
    \begin{itemize}
        \item \( I_{ij} \): Importance rating on a scale of 1 (Not Important) to 5 (Very Important).
        \item \( C_{ij} \): Confidence rating on a scale of 1 (Very Uncertain) to 5 (Very Certain).
    \end{itemize}
   \item Calculating Prior Mean \( \mu_{\omega_j} \): 
   The mean importance score for indicator \( j \), calculated across all respondents, represents \( \mu_{\omega_j} \):
   \[
   \mu_{\omega_j} = \frac{1}{n} \sum_{i=1}^n I_{ij}
   \]
   where \( n \) is the number of respondents.
   \item  Calculating Prior Variance \( \sigma_{\omega_j}^2 \):  
   The variance \( \sigma_{\omega_j}^2 \) depends on the mean confidence score, with a higher confidence resulting in a lower variance. Hence, the mean confidence score:
   \[
   \text{Mean Confidence}_j = \frac{1}{n} \sum_{i=1}^n C_{ij}
   \]
   will map the confidence score to the variance:
   \[
   \sigma_{\omega_j}^2 = \frac{\alpha}{\text{Mean Confidence}_j}
   \]
   where \( \alpha \) is a scaling factor.
   \item Gaussian Prior for Each Weight: The resulting Gaussian prior for each weight \( \omega_j \) is:
   \[
   \omega_j \sim \mathcal{N}(\mu_{\omega_j}, \sigma_{\omega_j}^2)
   \]
   This prior distribution reflects both the central tendency (importance) and dispersion (confidence) in the subjective beliefs of children and mothers about each indicator.
\end{enumerate}

\subsection{Likelihood Function and Posterior estimates}

Given data matrix \( X \in \mathbb{R}^{n \times k} \) with \( n \) children and \( k \) binary indicators, the likelihood and prior are defined as follows:

1. \textbf{Bernoulli likelihood per Indicator.} 
   For each indicator \( j \) and child \( i \), we model the probability of \( x_{ij} \) given weight \( \omega_j \) as:
   \[
   p(x_{ij} | \omega_j) = \sigma(\omega_j)^{x_{ij}} (1 - \sigma(\omega_j))^{1 - x_{ij}}
   \]
   where \( \sigma(\omega_j) = \frac{1}{1 + e^{-\omega_j}} \).

2. \textbf{Joint likelihood}.  
   Assuming independence across children, the likelihood of observing \( X \) given weights \( \omega \) is:
   \[
   p(X | \omega) = \prod_{i=1}^n \prod_{j=1}^k \sigma(\omega_j)^{x_{ij}} (1 - \sigma(\omega_j))^{1 - x_{ij}}
   \]

3. \textbf{Gaussian prior on weights.}  
   With a Gaussian prior for each weight \( \omega_j \):
   \[
   \omega_j \sim \mathcal{N}(\mu_{\omega_j}, \sigma_{\omega_j}^2)
   \]
   the joint prior is:
   \[
   p(\omega) = \prod_{j=1}^k \mathcal{N}(\omega_j | \mu_{\omega_j}, \sigma_{\omega_j}^2)
   \]

4. \textbf{Posterior approximation. } 
   The posterior for \( \omega \) given \( X \) is approximated as a Gaussian with mean and variance:
   \[
   \mu_{\omega_j}' = \frac{\sigma_{\omega_j}^2 \sum_{i=1}^n (x_{ij} - \sigma(\omega_j)) + \mu_{\omega_j} \sigma_j^2}{\sum_{i=1}^n x_{ij}^2 + \sigma_j^2}
   \]
   \[
   \sigma_{\omega_j}'^2 = \left( \frac{1}{\sigma_{\omega_j}^2} + \sum_{i=1}^n x_{ij}^2 \right)^{-1}
   \]
   leading to:
   \[
   p(\omega | X) \approx \mathcal{N}(\mu', \Sigma')
   \]
   where \( \mu   \mu' = (\mu_{\omega_1}', \mu_{\omega_2}', \ldots, \mu_{\omega_k}') \) and \( \Sigma' = \text{diag}(\sigma_{\omega_1}'^2, \sigma_{\omega_2}'^2, \ldots, \sigma_{\omega_k}'^2) \).

\section{Optimizing certainty-adjusted weights}\label{sec:ga}

In this process, we aim to develop a robust method for optimizing certainty-adjusted weights in a neural network model. The motivation stems from the need to accurately capture subjective responses, which vary not only in content (e.g., ratings on a Likert scale) but also in certainty, as inferred from response times captured by the mobile app. By using response time as a measure of (un)certainty, we can adjust each response, with quicker responses indicating greater confidence. This certainty adjustment allows us to weigh each response according to its reliability, leading to a more nuanced understanding of subjective data and thus obtain more precise digital phenotypes. 

\subsection{Certainty-Adjusted Likert Responses}
   For each response \( y_{ij} \) on a Likert scale from child or mother \( i \) for indicator \( j \), we define a certainty score \( c_{ij} \) based on the response time \( t_{ij} \):
   \[
   c_{ij} = \frac{1}{1 + \alpha \cdot t_{ij}}
   \]
   where \( \alpha \) is a scaling factor that adjusts the sensitivity of the certainty score to response time. The certainty-adjusted response \( y_{ij}^{(c)} \) then becomes:
   \[
   y_{ij}^{(c)} = y_{ij} \cdot c_{ij}
   \]

\subsection{Neural Network Model for the Fitness Function}
   Define a neural network \( f_{\psi}(\omega) \) with parameters \( \psi \) (synaptic weights and biases), where \( f_{\psi}(\omega) \) maps the weight vector \( \omega = (\omega_1, \omega_2, \ldots, \omega_k) \) to a predicted fitness score. This network approximates the posterior probability:
   \[
   f_{\psi}(\omega) \approx p(Y^{(c)} | \omega) \cdot p(\omega)
   \]
   where \( p(Y^{(c)} | \omega) \) is the likelihood of observing the certainty-adjusted responses \( Y^{(c)} \) given weights \( \omega \), and \( p(\omega) \) is the prior based on previous posteriors.

   Training the Neural Network:  
   The neural network is trained using an initial dataset of weight vectors \( \{\omega^{(i)}\}_{i=1}^N \) and their corresponding true fitness values, computed as \( \text{TrueFitness}(\omega^{(i)}) = p(Y^{(c)} | \omega^{(i)}) \cdot p(\omega^{(i)}) \). We minimize the following loss function:
   \[
   \text{Loss}(\psi) = \frac{1}{N} \sum_{i=1}^N \left( f_{\psi}(\omega^{(i)}) - \text{TrueFitness}(\omega^{(i)}) \right)^2
   \]
   where \( \psi \) denotes the neural network parameters (synaptic weights and biases), \( N \) is the size of the training set, and \( f_{\psi}(\omega^{(i)}) \) is the neural network's predicted fitness for input \( \omega^{(i)} \).

To explicitly define the loss function used to train the neural network \( f_{\psi}(\omega) \), which predicts the fitness score based on certainty-adjusted weights \( \omega \), we proceed as follows:

Given:
- \( \psi \) as the vector of neural network parameters (synaptic weights and biases),
- \( f_{\psi}(\omega^{(i)}) \) as the neural network’s predicted fitness for the weight vector \( \omega^{(i)} \),
- \( \text{TrueFitness}(\omega^{(i)}) \) as the true fitness score for \( \omega^{(i)} \), defined as the product of the likelihood \( p(Y^{(c)} | \omega^{(i)}) \) and prior \( p(\omega^{(i)}) \),

we aim to minimize the error between the network’s predicted fitness and the actual fitness across a training set of size \( N \).

Mathematical Definition of the Loss Function

The loss function \( \text{Loss}(\psi) \) is defined as the Mean Squared Error (MSE) between the predicted fitness values and the true fitness values for each sample \( \omega^{(i)} \) in the training set:

\[
\text{Loss}(\psi) = \frac{1}{N} \sum_{i=1}^N \left( f_{\psi}(\omega^{(i)}) - \text{TrueFitness}(\omega^{(i)}) \right)^2
\]

where \( N \) is the number of samples in the training set, and \( f_{\psi}(\omega^{(i)}) \) is the neural network’s prediction of the fitness score for the weight vector \( \omega^{(i)} \), and 
\[ \text{TrueFitness}(\omega^{(i)}) = p(Y^{(c)} | \omega^{(i)}) \cdot p(\omega^{(i)}) \] combines the certainty-adjusted likelihood \( p(Y^{(c)} | \omega^{(i)}) \) and the prior \( p(\omega^{(i)}) \).

To explicitly define \( \text{TrueFitness}(\omega^{(i)}) \), we can use the certainty-adjusted responses in the likelihood and assume a Gaussian prior on each weight \( \omega_j^{(i)} \) to obtain a certainty-adjusted likelihood \( p(Y^{(c)} | \omega^{(i)}) \):
   \[
   p(Y^{(c)} | \omega^{(i)}) = \prod_{j=1}^k \mathcal{N}(y_{ij}^{(c)} | \omega_j^{(i)}, \tau^2)
   \]
   where \( y_{ij}^{(c)} = y_{ij} \cdot c_{ij} \) is the certainty-adjusted response, and \( \tau^2 \) is the variance of the likelihood.

For the prior \( p(\omega^{(i)}) \), assume a Gaussian prior on each weight \( \omega_j^{(i)} \):
   \[
   p(\omega^{(i)}) = \prod_{j=1}^k \mathcal{N}(\omega_j^{(i)} | \mu_j, \sigma_j^2)
   \]
Thus, the full expression for \( \text{TrueFitness}(\omega^{(i)}) \) becomes:
\[
\text{TrueFitness}(\omega^{(i)}) = \left( \prod_{j=1}^k \mathcal{N}(y_{ij}^{(c)} | \omega_j^{(i)}, \tau^2) \right) \cdot \left( \prod_{j=1}^k \mathcal{N}(\omega_j^{(i)} | \mu_j, \sigma_j^2) \right)
\]

Substituting \( \text{TrueFitness}(\omega^{(i)}) \) back into the loss function:
\[
\text{Loss}(\psi) = \frac{1}{N} \sum_{i=1}^N \left( f_{\psi}(\omega^{(i)}) - \left( \prod_{j=1}^k \mathcal{N}(y_{ij}^{(c)} | \omega_j^{(i)}, \tau^2) \cdot \prod_{j=1}^k \mathcal{N}(\omega_j^{(i)} | \mu_j, \sigma_j^2) \right) \right)^2
\]

This function penalizes deviations between the predicted fitness and the true fitness, based on the certainty-adjusted responses and prior beliefs, and trains the neural network to approximate the posterior likelihood as accurately as possible.

\subsection{Optimizing the Neural Network Synapses with Genetic Algorithms}
   To optimize the neural network’s synaptic parameters \( \psi \), we use a genetic algorithm to evolve the network’s weights and biases. This evolutionary optimization proceeds through the following steps:

\begin{itemize}
    \item Encoding Neural Network Parameters:  
     Each individual in the population represents a complete set of network parameters \( \psi = (\psi_1, \psi_2, \ldots, \psi_n) \), where \( \psi_i \) are the synaptic weights and biases. These parameters are flattened into a vector and form the "chromosome" for each individual in the genetic algorithm.
     \item Fitness Evaluation:  
     For each individual (network configuration) \( \psi^{(j)} \), compute the fitness as the negative loss function:
     \[
     \text{Fitness}(\psi^{(j)}) = -\text{Loss}(\psi^{(j)})
     \]
     This negative loss is maximized in the genetic algorithm, pushing the neural network parameters toward configurations that minimize the original loss function.
     \item Selection: Select individuals based on fitness, with preference given to those achieving a lower loss (higher fitness). Standard selection techniques, such as tournament selection or roulette wheel selection, are used to identify the best-performing individuals.
     \item Crossover: Perform crossover between selected individuals by combining sections of their parameter vectors \( \psi \). For example, given two parent vectors \( \psi^{(p1)} \) and \( \psi^{(p2)} \), a new offspring vector \( \psi^{(c)} \) is created by:
     \[
     \psi^{(c)} = \alpha \cdot \psi^{(p1)} + (1 - \alpha) \cdot \psi^{(p2)}
     \]
     where \( \alpha \in [0, 1] \) is a crossover rate. This operation allows offspring to inherit characteristics from both parents, enabling exploration of new parameter combinations.
     \item Mutation: To introduce diversity and explore additional regions of the parameter space, apply mutation to the offspring vectors. Each parameter \( \psi_i \) in \( \psi^{(c)} \) undergoes random perturbation:
     \[
     \psi_i = \psi_i + \epsilon, \quad \epsilon \sim \mathcal{N}(0, \sigma_{\text{mutate}}^2)
     \]
     where \( \epsilon \) is drawn from a Gaussian distribution with variance \( \sigma_{\text{mutate}}^2 \). The mutation rate and variance are controlled to balance exploration and convergence.
     \item Iteration and Convergence: Repeat the selection, crossover, and mutation steps over multiple generations. The genetic algorithm iteratively refines the neural network parameters \( \psi \), evolving them toward configurations that minimize the loss function. Convergence is achieved when the population stabilizes, or when the fitness scores reach a satisfactory level.
\end{itemize}

The previous approach leverages the genetic algorithm to optimize both the certainty-adjusted weights and the synaptic weights of the neural network model used to evaluate fitness. By encoding the neural network’s synaptic parameters as chromosomes, the GA iteratively evolves the network toward configurations that accurately capture the relationship between certainty-adjusted responses and their fitness. This dual-level optimization enables the neural network model to better represent complex, certainty-weighted subjective data, achieving a robust fit that is not only reflective of individual confidence levels but also adaptable through evolutionary refinement.

\section{Conclusion}\label{sec:con}
This paper presents a novel Bayesian artificial intelligence framework for constructing a Multidimensional Index of Child Growth (MICG) using digital phenotyping. By embedding this algorithm within a mobile application, we offer a scalable solution for tracking child development across multiple dimensions, addressing the limitations of traditional, physically focused growth models. The Bayesian approach allows the model to dynamically incorporate new data from both objective indicators and subjective opinions gathered from mothers and children, capturing a broader picture of child well-being. Additionally, digital phenotyping techniques enable the app to weight responses according to inferred certainty levels based on response times, thus enhancing the model’s accuracy. This framework represents a significant advancement in continuous and multidimensional child monitoring, offering a practical, real-time tool for families, healthcare providers, and researchers to support early childhood development in a holistic manner. Future work will explore the longitudinal impacts of using MICG to predict developmental outcomes and assess the broader applications of digital phenotyping in public health.

\bibliographystyle{plainnat}
\bibliography{references}

\end{document}